\documentstyle[epsf,12pt]{article}
\title{\vspace*{-3cm}\hfill {\small BUTP-95/13}\hspace{2cm} \\[3cm]
The $m^4 \ln m\,$ Contribution \\ to the Nucleon Mass in CHPT\\[10mm]}
\author{A. Kallen\\[5mm]
Institut f\"ur Theoretische Physik, Universit\"at Bern \\
Sidlerstrasse 5, CH-3012 Bern, Switzerland\\[4mm]} 
\date{\ifcase\month\or January\or February\or March\or April\or May\or
June\or July\or August\or September\or October\or November\or December\fi
\space\number\year}

\begin{document}
\maketitle
\vspace*{\fill}
\begin{center}
{\bf Abstract}

\vspace*{8mm}

\begin{minipage}{13cm} 
In CHPT the hadron masses obtain loop corrections from the pseudoscalar
mesons which are identified with the Goldstone bosons of broken chiral
symmetry. An expansion of the baryon mass in the quark masses therefore
includes non-analytic terms. We calculate the nucleon mass in the one-loop
approximation to order $m^4 \sim m_q^2$. We compare the
result in the relativistic and in the heavy mass formulation of the
theory and derive matching relations between corresponding low-energy
constants. We calculate the pion-nucleon loop in old-fashioned perturbation
theory and find a contribution to the $m^4 \ln m$ term from an
intermediate state which does not occur in the heavy baryon theory.
\end{minipage}
\end{center}
\vspace*{\fill}
\thispagestyle{empty}
\pagebreak

   \section{Introduction}

The QCD Lagrangian for $n$ flavours of massless quarks is invariant
under chiral rotations of the fields. The spontaneous breaking of the
$SU(n)_R \times SU(n)_L$ chiral symmetry to $SU(n)_V$ gives rise to
$n^2-1$ Goldstone particles which lead to infrared divergences. In the
presence of small quark masses the Goldstone bosons also become massive
and are identified with the light pseudoscalar mesons. The square of the
meson mass $m^2$ is proportional to the quark mass $m_q$. The mass of a
hadronic bound state is determined by the QCD scale and by the quark
masses. Due to the Goldstone nature of the mesons the expansion of the
hadron mass in the quark masses is not a simple Taylor series but
contains non-analytic terms proportional to $m_q^{3/2}$ and $m_q^2 \ln
m_q$.
 
Hadronic low-energy processes cannot be computed directly from the QCD
Lagrangian but they can be described by an effective Lagrangian which
reproduces the correct symmetry properties. Chiral Perturbation Theory
(CHPT) is based on a simultaneous expansion in powers of derivatives and
of quark masses~\cite{GasserLeutwyler}. In a relativistic formulation of
the effective theory the loop expansion coincides with the expansion in
small momenta and masses. In the mesonic sector higher dimension
operators are typically suppressed by factors of $m^2/\Lambda^2$ where
$\Lambda = 4 \pi F_\pi$ is the QCD scale of order 1 GeV.

In the baryonic sector the chiral expansion is complicated by the
additional scale introduced by the mass $M$ in the baryon propagator
which is of the same order as $\Lambda$ itself.  The loop expansion no
longer coincides with the small momentum expansion. A consistent power
counting is only possible if the baryon kinematics is treated in a
non-relativistic framework~\cite{Jenkins}. 

In this article we consider the dependence of the nucleon mass on the
masses of the two lightest quark flavours, the $u$ and $d$ quark,
keeping $m_s$ fixed. In the two-flavour sector the low-energy particle
spectrum consists of nucleons and pions. We work in the isospin limit,
setting $m_u = m_d$. Then the proton and neutron form a mass degenerate
isospin doublet while the pions belong to a triplet. The expansion of
the pion mass squared starts with $m_\pi^2 = (m_u + m_d)B + O(m_q^2 \ln
m_q)$ where $B$ is related to the quark condensate. It is convenient to
use the abbreviation $m^2 \equiv (m_u + m_d)B$. The expansion of the
nucleon mass is given by
\begin{equation}
M_{phys.} = a_0 + a_1 m^2 - \frac{3g_A^2}{32\pi F^2} m^3 + a_2 m^4 \ln
\frac{m^2}{\mu^2} + a_3 m^4 + O(m^5)\ . 
\label{e:mass} 
\end{equation}
$a_0$ is the value of the nucleon mass in the chiral limit where $m=0$.
The coefficient $a_2$ receives contributions from pion-nucleon loops and
pion tadpoles. In this article we compare these contributions in CHPT
and in the heavy baryon expansion. This comparison leads to matching
conditions for corresponding low-energy constants. With the help of
non-relativistic old-fashioned perturbation theory we can then show that
the pion-nucleon loop contribution to $a_2$ in CHPT is due to two
different intermediate states. One of them contains an anti-nucleon and
has no corresponding counterpart in the heavy baryon theory. 

This article is organized in the following way. In section
\ref{Lagrangian} we list those terms in the pion-nucleon  Lagrangian
which are relevant for the calculation of the nucleon mass. We give them
in the relativistic theory and in the heavy baryon approximation.
Section \ref{Shift} contains the results for the expansion coefficients
$a_i$ in both approaches and matching conditions for the corresponding
low-energy constants. In section \ref{Origin} we calculate the
pion-nucleon loop contribution in old-fashioned perturbation theory and
discuss the origin of the non-analytic terms, in particular of $m^4 \ln
m$, in the different approaches. Section \ref{Summary} contains a short
summary of the results.

   \section{The Effective Lagrangian} \label{Lagrangian} 

The chiral effective Lagrangian for baryons has been given by
Gasser~et~al.~\cite{GasserSainio}. It is given by a string of
interaction Lagrangians of increasing chiral power,
\begin{displaymath} 
{\cal L}_{\pi N}\ =\ {\cal L}^{(1)} \ +\ {\cal L}^{(2)} \ +\ {\cal
L}^{(3)} \ +\ \cdots \ .
\end{displaymath} 
The superscript $(i)$ in ${\cal L}^{(i)}$ gives the number of
derivatives and/or powers of quark mass. ${\cal L}^{(1)}$ contains the
free nucleon Lagrangian.  The nucleon mass to order $m^4$ in the
one-loop approximation is determined by tree graphs and one-loop graphs
with vertices from ${\cal L}^{(1)}$ and ${\cal L}^{(2)}$. From the
higher order Lagrangians ${\cal L}^{(3)}$ and ${\cal L}^{(4)}$ we only
have to consider tree diagrams. In the calculation of the self-energy
there arises a term of dimension zero which requires the introduction of
a counterterm Lagrangian ${\cal L}^{(0)}$. Explicitly we have  
\begin{eqnarray*} 
{\cal L}^{(0)} &=& \Delta M\ \bar{\psi} \psi\ , \\
{\cal L}^{(1)} &=& \frac{g_A}{2}\ \bar{\psi}\ {u\!\!\!/}
\gamma_5\ \psi \ +\ \cdots \ , \\ 
{\cal L}^{(2)} &=& \frac{M}{F^2}\  \bar{\psi} \left[ c_1\, m^2
<U^\dagger + U >\ -\ \frac{c_3}{4} <u \cdot u > \right] \psi \ +\ 
\cdots\ , \\
{\cal L}^{(4)} &=&  d\, m^4\ \bar{\psi} \psi \ +\ \cdots \ .
\end{eqnarray*} 
Brackets denote the trace in isospin space, $M$ is the bare nucleon
mass, $F\simeq 93$~MeV the pion decay constant and $g_A\simeq 1.25$ is
the axial-vector coupling constant. The three pions are parametrized
with the help of the Pauli matrices $\tau$,
\begin{eqnarray*}
U &=& u^2 \ =\ \exp \frac{i}{F}\,  \vec{\pi} \cdot \vec{\tau}\ , \\ 
u_\mu &=& i u^\dagger\, \partial_\mu U\, u^\dagger \ .
\end{eqnarray*} 
There is no term in ${\cal L}^{(3)}$ which can give a contribution to
the nucleon mass up to order $m^4$. The term proportional to $c_2$ in
${\cal L}^{(2)}$ does not contribute to the nucleon mass. Thus we have
omitted it, together with other interaction terms ($ + \cdots$) which
are irrelevant for the determination of the coefficients $a_2$ and $a_3$
of equation (\ref{e:mass}). The constants $c_1,\, c_3\,$ and $d$ are
divergent and have to be renormalized. 
\begin{eqnarray*}
&&c_i \ =\ c_i^R(\mu) + \gamma_i\, L \hspace*{5mm} (i=1,3) \hspace*{15mm} 
\gamma_1 = - \frac{3g_A^2}{4}\ , \hspace*{8mm} \gamma_3 = 4 g_A^2 \\
&&d \ =\ d^R(\mu) + \delta\, L \hspace*{32mm}  \delta = 4 c_1^R(\mu) +
c_3^R(\mu) \\
&&L\ =\ \frac{\mu^{d-4}}{16\pi^2}\left[ \frac{1}{d-4} - \frac{1}{2} \left(
\ln 4\pi + \Gamma^\prime (1) + 1 \right) \right] 
\end{eqnarray*}
$\mu$ is the renormalization scale introduced by dimensional
regularization. The finite low-energy constants $c_i^R(\mu)$ are a
priori undetermined but can in principle be fixed from phenomenology.
Their corresponding counterparts $c_i^\prime$ in the heavy baryon
approximation have been calculated (see equation (\ref{e:lecs}) below).
In section \ref{Shift} we derive matching conditions between the
low-energy constants of both theories at the renormalization scale $\mu
= M$ and give numerical values for $c_1^R(M)$ and $c_3^R(M)$ (equation
(\ref{e:mc}) in section \ref{Shift}).

For an introduction to heavy baryon chiral perturbation theory, see e.g.
refs.~\cite{Jenkins,Bernard}. The nucleon momentum is written as $p^\mu
= M v^\mu + k^\mu$ with $v^2 = 1$ and $v\cdot k \ll M$. After
decomposing the wave-function $\psi$ into two components $B$ and $b$ and
integrating out the heavy field $b$ we are left with the
velocity-dependent baryon field  
\begin{displaymath} 
B(v,x)=\exp\, ( i M v\cdot x )\ \frac{(1+ {v\!\!\!/})}{2}\ \psi (x)
\ .
\end{displaymath} 
This redefinition transforms the free nucleon Lagrangian for the massive
field $\psi$ into a free Lagrangian for $B$ plus a string of terms which
are suppressed by factors of $1/M$. The Dirac equation for the massless
field $B$ to leading order is given by 
\begin{displaymath}
i v \cdot \partial B = 0 \ .
\end{displaymath}

Explicitly the relevant interaction terms for the calculation of the
nucleon mass are
\begin{eqnarray*}
{\cal L}_{hb}^{(1)} &=& g_A\ \overline{B}\ S \cdot u\ B\ , \\
{\cal L}_{hb}^{(2)} &=& \frac{1}{2M}\ \overline{B} \left[ (v \cdot
\partial)^2 \, -\, \partial^2\, -\, i g_A \left\{ S\cdot \partial\, ,\,
v \cdot u \right\}\, +\, c^\prime_1 m^2 <U^\dagger + U > \, + \right. \\
&&\left. \hspace*{15mm} \, +\, \left( c^\prime_2 - \frac{g_A^2}{2}
\right) <(v\cdot u)^2 >\, +\, \left( c^\prime_3 + \frac{g_A^2}{4}
\right) < u \cdot u >\, \right] B \ , \\ 
{\cal L}_{hb}^{(4)} &=& \frac{d^\prime}{M}\ m^4\ \overline{B} B\ .
\end{eqnarray*}
Note that the Dirac matrices $\gamma^\mu$ have been expressed in terms
of $v^\mu$ and the velocity-dependent spin operator $S^\mu =
\frac{i}{2}\gamma_5 \sigma^{\mu\nu} v_\nu$ which obeys $S \cdot v = 0$.
In the heavy baryon approximation the nucleon mass vanishes in the
chiral limit, therefore there is no ${\cal L}_{hb}^{(0)}$. The terms
in ${\cal L}_{hb}^{(2)}$ with have fixed coefficients are due to the
$1/M$ expansion of the Lagrangian ${\cal L}^{(1)}$. The first two
contributions are corrections to the kinetic energy. The term
proportional to $(v \cdot \partial)^2$ does not contribute to matrix
elements at order $1/M$ because of the leading order Dirac equation for
$B$. The counterterm $d^\prime$ has to be renormalized.
\begin{displaymath}
d^\prime \ =\ d^{\prime R}(\mu) + \delta^\prime \, L \hspace*{15mm}
\delta^\prime = \frac{1}{2 F^2}\, \left( - c^\prime_1 + \frac{1}{4}
c^\prime_2 + c^\prime_3 - \frac{1}{16} g_A^2 \right)
\end{displaymath} 

The low-energy coefficients $c_i^\prime$ are finite. They are determined
from the pion-nucleon $\sigma$-term and from $\pi N$ scattering lengths.
We take the numerical values given by Mei\ss ner in ref.~\cite{Meissner}
where one can find a critical discussion of these numbers and their
errors.
\begin{equation}
\begin{array}{lll}
c^\prime_1 = -1.63 \pm 0.21 \hspace*{10mm} & \sigma_{\pi N} &
\cite{Bernard} \\[4mm] 
c^\prime_2 = 6.20 \pm 0.38 & \pi N \rightarrow \pi N & \cite{BKM} \\[4mm]
c^\prime_3 = -9.86 \pm 0.41 & \pi N \rightarrow \pi N & \cite{BKM}
\end{array} 
\label{e:lecs}
\end{equation}

\vfill

   \section{The Nucleon Mass} \label{Shift}

The relativistic calculation yields the following results for
the coefficients $a_i$ of equation (\ref{e:mass})
\begin{eqnarray*}
a_0 &=& M\ , \\[2mm]
a_1 &=& \frac{M}{F^2} \left[ \frac{3g_A^2}{32\pi^2} \left( 1 - \ln
\frac{M^2}{\mu^2} \right) - 4 c^R_1(\mu) \right]\ , \\[2mm]
a_2 &=& \frac{1}{32\pi^2 F^2} \left[ \frac{M}{F^2} \left( 4 c_1^R(\mu) +
c_3^R(\mu) \right) - \frac{3 g_A^2}{2 M} \right]\ , \\[2mm]
a_3 &=& \frac{3g_A^2}{32\pi^2 F^2 M} \left( 1 + \frac{1}{2} \ln
\frac{M^2}{\mu^2} \right) - d^R(\mu)\ .
\end{eqnarray*}
We have chosen the value of $\Delta M$ in such a way that $M_{phys.} =
M = a_0$ in the chiral limit. The result in the heavy baryon limit is 
\begin{eqnarray*}
a_0 &=& 0\ , \\[2mm]
a_1 &=& - \frac{2}{M}\, c^\prime_1\ , \\[2mm]
a_2 &=& \frac{1}{32\pi^2 F^2}\ \frac{1}{2M} \left( - c^\prime_1 +
\frac{1}{4} c^\prime_2 + c^\prime_3 - \frac{1}{16} g_A^2 \right)\ , \\[2mm]
a_3 &=& - \frac{1}{M} d^{\prime R}(\mu)\ .
\end{eqnarray*}

Comparing the results for the $a_i$ of the two theories leads to the
following matching conditions at the renormalization scale $\mu = M$

\newpage

\begin{equation}
\begin{array}{l}
c_1^R (M) = {\displaystyle \frac{F^2}{2 M^2}}\, c^\prime_1 \ +\
{\displaystyle \frac{3g_A^2}{128\pi^2} } \ , \\[4mm]
c_3^R (M) = {\displaystyle \frac{F^2}{2 M^2}} \left( -5 c^\prime_1 +
\frac{1}{4}  c^\prime_2 +  c^\prime_3 \right) + g_A^2 \left(
{\displaystyle \frac{F^2}{M^2} \frac{47}{32} - \frac{3}{32\pi^2}} \right) \
, \\[4mm] 
d^R (M) = {\displaystyle \frac{1}{M}}\left( d^{\prime R}(M) +
{\displaystyle \frac{3g_A^2}{32\pi^2 F^2}} \right) \ .
\end{array}
\end{equation}
Note that the condition for $c_1^R (M)$ is the one given by
Bernard~et~al. in ref.~\cite{Bernard}. With the values of equation
(\ref{e:lecs}) for the $c_i^\prime$ and taking $F= 93$~MeV, $M= 940$~MeV
and $g_A = 1.25$ we obtain the numerical results 
\begin{equation}
\begin{array}{l} 
c_1^R (M) = ( - 4.27 \pm 1.03 ) \cdot 10^{-3}\ , \\[4mm]
c_3^R (M) = ( 6.84 \pm 7.61 ) \cdot 10^{-3}\ .
\end{array}  
\label{e:mc}
\end{equation} 

The huge error of $c_3^R (M)$ is due to the fact that the central values
almost cancel while the errors sum up: $(-5 c_1^\prime + \frac{1}{4}
c_2^\prime + c_3^\prime ) = ( -0.16 \pm 1.56)$.

   \section{The Origin of the $m^4\ln m\,$ term} \label{Origin}

The coefficient $a_2$ of the non-analytic term proportional to $m^4\,
\ln m$ receives con\-tribu\-tions both from the pion-\-nucleon loop and from
pion tad\-poles. How\-ever, depending on which theory one uses to calculate
the nucleon mass, the loop and the counterterms play a different role.
The coefficient of the leading non-analytic term proportional to $m^3$
on the other hand is entirely due to the pion-nucleon loop in both
models. In the relativistic theory we have
\begin{eqnarray*}
&&a_{2, loop} = \frac{3 g_A^2}{32\pi^2 F^2 M} \left( -
\frac{1}{2} \right)\ , \\[2mm]
&&a_{2, tadpole} = \frac{1}{32\pi^2 F^2}\ \frac{M}{F^2} \left( 4
c_1^R(\mu) + c_3^R(\mu) \right)\ .
\end{eqnarray*}

We can calculate $a_{2, loop}$ from the pion-nucleon interaction in
${\cal L}^{(1)}$ by using old-fashioned perturbation theory. There one
looks for a solution of the eigenvalue problem 
\begin{displaymath}
(H_0 + g_A H_1) |\phi\rangle = \left( \sum_n g_A^n E_n \right)
|\phi\rangle \ .
\end{displaymath}
To second order ($n = 2$) the solution is given by the standard formula
\begin{displaymath}
E\ =\ E_0\ +\ g_A^2\, \sum_{int.}\ \frac{\left|\, \langle int. | H_1 |
\phi_0 \rangle \, \right|^2}{E_0 - E(int.)}\ .
\end{displaymath}
where $H_0, E_0$ and $\phi_0$ are the unperturbed quantities and
$E(int.)$ is the energy of the intermediate state $|int. \rangle$. The
energy-shift directly translates into the mass-shift and we can
therefore investigate the contributions of the possible intermediate
states to the coefficients $a_i$. $H_1$ is proportional to ${\cal
L}^{(1)}$ and we find two intermediate states which give a non-vanishing
matrix element. They are depicted in figure \ref{f:}.  State (a)
contains one pion and one nucleon, state (b) contains one pion, two
nucleons and one anti-nucleon. Their matrix elements are determined by
three-dimensional integrals,
\begin{eqnarray*}
\frac{\left|\, \langle (a) | H_1 | \phi_0 \rangle \,
\right|^2}{E_0 - E(a)} & \leftrightarrow & \int\!\!
\frac{d^{3}l}{(2\pi)^{3}}\ \frac{1}{e(m)\, e(M)}\ \frac{1}{M - e(m) -
e(M)}\ , \\[2mm]
\frac{\left|\, \langle (b) | H_1 | \phi_0 \rangle \,
\right|^2}{E_0 - E(b)} &\ \leftrightarrow\ & \int\!\!
\frac{d^{3}l}{(2\pi)^{3}}\ \frac{1}{e(m)\, e(M)}\ \frac{(-1)}{M +
e(m) + e(M)}\ , \\[2mm]
e(x) = \sqrt{x^2 + \vec{l}^{\ 2}\,}\ .
\end{eqnarray*}

The sum of the two contributions is just the relativistic loop integral
in the rest frame of the nucleon, $p = (M, \vec{0}\,)$, after
integration over the time component $l^0$.
\begin{eqnarray*}
\lefteqn{\int\!\! \frac{d^{4}l}{(2\pi)^{4}}\ \frac{1}{m^2 - l^2}\
\frac{1}{M^2 - (p-l)^2}\ \sim} \\[2mm]
&&\ \int\!\! \frac{d^{3}l}{(2\pi)^{3}}\ \frac{1}{e(m)\, e(M)}\
\left[ \frac{1}{M - e(m) - e(M)}\ -\ \frac{1}{M + e(m)
+ e(M)}\right] 
\end{eqnarray*}

We expand the integrands in terms of $m$ and use dimensional
regularization to evaluate the integrals. The leading non-analytic term
proportional to $m^3$ in equation (\ref{e:mass}) is entirely due to the
intermediate state (a) as we would expect. The contributions to the
coefficient $a_{2, loop} = a_2^{(a)} + a_2^{(b)}$ from the two states
are given by
\begin{displaymath}
\begin{array}{l}
a_2^{(a)} = {\displaystyle \frac{3g_A^2}{32\pi^2 F^2 M} \left( - \frac{3}{4}
\right)}\ , \\[4mm]
a_2^{(b)} = {\displaystyle \frac{3g_A^2}{32\pi^2 F^2 M} \left( \frac{1}{4}
\right)}\ .
\end{array}
\end{displaymath}

The reason that state (a) contributes to the logarithmic divergence is
clearly due to the singular behaviour of the denominator in the
integral: $[M - e(m) - e(M)]$ is infrared divergent for vanishing $m^2$
(chiral limit). Still it does not give the full contribution to $a_2$.

The surprising result is, that there is also a non-zero $a_2^{(b)}$. The
integral which gives the contribution from state (b) has a denominator
proportional to $2M$ in the chiral limit, it is finite. Obviously, its
derivative with respect to $m^2$ is not and the singularity comes from
the expansion of the integrand in terms of $m$. The intermediate state
(b) contains an anti-particle, a configuration which does not occur in
the framework of the heavy baryon expansion.  Here the loop contribution
from ${\cal L}_{hb}^{(1)} \times {\cal L}_{hb}^{(1)}$ is independent of
$M$. It gives just the non-analytic term proportional to $m^3$ and no
contribution to $a_2$ or $a_3$. The non-zero contribution to $a_2$ comes
from the ${\cal L}_{hb}^{(1)} \times {\cal L}_{hb}^{(2)}$ loop. We have 
\begin{eqnarray*}
&&a_{2, loop} = \frac{3 g_A^2}{32\pi^2 F^2 M} \left( - \frac{1}{32}
\right)\ , \\[2mm]
&&a_{2, tadpole} = \frac{1}{32\pi^2 F^2}\ \frac{1}{2M} \left( -
c^\prime_1 + \frac{1}{4} c^\prime_2 + c^\prime_3 + \frac{1}{8} g_A^2
\right)\ .
\end{eqnarray*}

For completeness we also give the results for the coefficient $a_3$. In
the relativistic calculation we have $a_{3, loop} = a_3^{(a)} +
a_3^{(b)}$ with 
\begin{displaymath}
\begin{array}{l}
a_3^{(a)} = {\displaystyle \frac{3g_A^2}{32\pi^2 F^2 M} \left( -
\frac{1}{4} + 2 \ln 2 + \frac{3}{4} \ln \frac{M^2}{\mu^2} \right)}\ ,
\\[4mm]
a_3^{(b)} = {\displaystyle \frac{3g_A^2}{32\pi^2 F^2 M} \left(
\frac{5}{4} - 2 \ln 2 - \frac{1}{4} \ln \frac{M^2}{\mu^2} \right)}\ ,
\end{array}
\end{displaymath}
while the heavy baryon calculation gives $a_{3, loop} = 0$.

We have the following results. The pion-nucleon loop is always
responsible for the leading non-analytic term proportional to $m^3$. In
the relativistic theory, the loop of ${\cal L}^{(1)} \times {\cal
L}^{(1)}$ contributes to the coefficients $a_2$ and $a_3$. In
old-fashioned perturbation theory it is the intermediate state (a) with
one meson and one baryon which gives the leading singularity. The
perturbative treatment shows that state (b), which contains an
anti-baryon, gives also a contribution to $a_2$. We would not have
expected this, since the integral expression which describes its
contribution to the nucleon mass looks perfectly innocuous. In the heavy
mass theory the loop of ${\cal L}^{(1)}_{hb} \times {\cal
L}^{(1)}_{hb}$ can only contribute to order $m^3$. It gives no
contribution to $a_2$ or $a_3$. The $m^4 \ln m\,$ terms are due to pion
tadpole contributions of ${\cal L}^{(2)}_{hb}$ and to the loop of ${\cal
L}^{(1)}_{hb} \times {\cal L}^{(2)}_{hb}$, which are both suppressed by
a factor of $1/M$.

   \section{Summary} \label{Summary}

We have calculated the nucleon mass in the one-loop approximation to order
$m^4 \sim m_q^2$ in CHPT and in the heavy baryon approximation. We
reproduce the matching condition for the low-energy constant $c_1^R (M)$ of
Bernard et~al. given in~\cite{Bernard} and give similar conditions for
$c_3^R (M)$ and the counter\-term $d$. We have used old-fashioned
perturbation theory to decompose the contribution of the relativistic
pion-nucleon loop into the sum of terms corresponding to two different
intermediate loop states. The leading singularity proportional to $m^3$ is
always due to the pion-nucleon loop. The coefficient $a_2$ of the $m^4 \ln
m$ term in equation (\ref{e:mass}) can have contributions from the
pion-nucleon loop and from pion tadpole diagrams. These contributions are
model dependent.

\begin{figure}[hb]
\begin{center}
\mbox{\epsfxsize=13cm\epsfbox{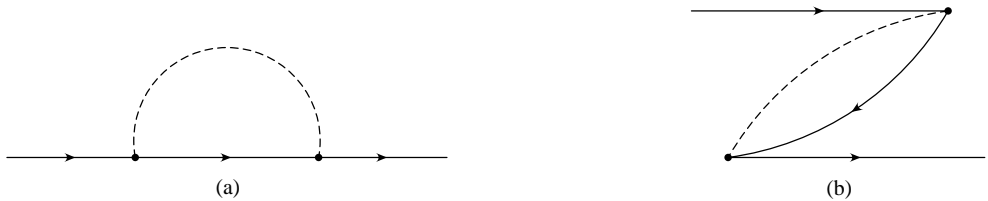}}
\end{center}
\caption{The two intermediate states which can contribute to the
nucleon mass to order $g_A^2$ in the perturbative approach. In (a) we
have one pion (dashed line) and one nucleon (solid line) while in (b) we
have three intermediate nucleon fields. Situation (b) is absent in the
heavy baryon approximation since anti-particles do not occur in that
framework.} \label{f:}
\end{figure}

\begin{itemize} 
\item In the relativistic theory there is a contribution to the
logarithmic divergence from the loop of ${\cal L}^{(1)} \times
{\cal L}^{(1)}$. Other contributions come from the tadpole diagrams of
${\cal L}^{(2)}$, they are proportional to the counterterms $c_1$ and
$c_3$. 
\item In the heavy baryon approximation the loop of ${\cal L}^{(1)}_{hb}
\times {\cal L}^{(1)}_{hb}$ gives no contribution to the logarithmic
term. To leading order in $1/M$ there is just the leading singularity
proportional to $m^3$ \cite{Bernard}. Consequently, all contributions to
$a_2$ are suppressed by a factor of $1/M$.

\item In old-fashioned non-relativistic perturbation theory there are
two intermediate states (see figure \ref{f:}). State (a) contains one
pion and one nucleon. The relevant integral gives an $m^4 \ln m\,$
contribution due to the vanishing of the denominator in the chiral
limit. State (b) contains one pion, two nucleons and one anti-nucleon
and also contributes to the $m^4 \ln m\,$ term although the denominator
in the corresponding integral is finite. In this case the logarithmic
singularity arises from the expansion of the integrand in $m$. 
\end{itemize}

I thank H. Leutwyler for helpful hints and discussions during this work.

\pagebreak

\end{document}